\documentclass[superscriptaddress,twocolumn,showpacs,prb,floatfix]{revtex4}
\usepackage{epsfig}
\usepackage{amssymb}

\begin{document}

\title{Size dependence of the internal energy in Ising and vector spin glasses}

\author{Helmut G.~Katzgraber}
\affiliation{Theoretische Physik, ETH H\"onggerberg,
CH-8093 Z\"urich, Switzerland}
\author{I.~A.~Campbell}
\affiliation{Laboratoire des Verres, Universit\'e Montpellier II, 34095
Montpellier, France}

\date{\today}

\begin{abstract}
We study numerically the scaling correction to the internal energy
per spin as a function of system size and temperature in a variety of
Ising and vector spin glasses. From a standard scaling analysis we
estimate the effective size correction exponent $x$ at each temperature. 
For each
system with a finite ordering temperature, as temperature is
increased from zero, $x$ initially decreases regularly until it goes
through a minimum at a temperature close to the critical
temperature, and then
increases strongly. The behavior of the exponent $x$ at and below the 
critical temperature
is more complex than suggested by the model for the size
correction that relates $x$ to the domain-wall stiffness exponent.

\end{abstract}

\maketitle

Since Gibbs, thermodynamic transitions have been classified according to
the critical behavior of the specific heat, or equivalently of the
critical temperature dependence of the internal energy. One of the
disturbing features of the spin-glass transition has always been that
there appears to be no thermodynamic signature whatsoever of the critical
temperature, except in the mean-field limit. In finite dimensions, the
specific-heat exponent $\alpha$ is strongly negative and the 
energy changes perfectly smoothly as a function of temperature through the
critical temperature $T_{\rm c}$.\cite{binder:86}

The standard scaling form for the finite-size correction to the internal
energy per spin $e(L)$ is
\begin{equation}
e(L) = e_\infty +aL^{-x} \; ,
\label{eq_energy}
\end{equation}
where $L$ represents the system size. An exponent $\theta_{\rm E}$
can be defined by $\theta_{\rm E}=d - x$ ($d$ represents the space
dimension). In Ising spin glasses (ISG) it has been surmised from
``droplet'' arguments\cite{newman:98,komori:99,bouchaud:02} that this
correction is directly related to the energy associated with domain
walls, for which independent numerical measurements can also be carried
out. Thus it is expected that at zero temperature $\theta_{\rm E}$ is
identical to $\theta_{\rm DW}$, the domain-wall stiffness exponent.  This
conjecture is related to the controversial question of the form of the
elementary excitations in spin glasses, and the identity should be valid
if periodic boundary conditions simply introduce supplementary domain
walls. It is exact for Migdal-Kadanoff spin glasses.\cite{bouchaud:02}
While the argument was introduced for zero temperature, it has also been
invoked for finite $T$.\cite{komori:99}
At a continuous transition the singular part of the free energy divided
by the temperature scales as ${\rm length}^{-d}$. Because 
$(T - T_{\rm c}) \sim {\rm length}^{-1/\nu}$ 
(see Refs.~\onlinecite{katzgraber:02a} and \onlinecite{barber:85}), 
if $T_{\rm c} > 0$,
\begin{equation}
x(T = T_{\rm c})= d - 1/\nu \; .
\label{eq_crit}
\end{equation}
If $T_{\rm c} = 0$, $\theta_{\rm DW} = -1/\nu$ at $T=0$  and
$x(0)= d + 1/\nu$.\cite{katzgraber:02a}

We have carried out Monte Carlo measurements of the size dependence of the
energy as a function of temperature in the mean-field
Sherrington-Kirkpatrick (SK) ISG model, in the Edwards-Anderson ISG with
Gaussian interactions in dimensions $2$, $3$, and $4$, in the gauge glass
(GG) in dimensions $2$, $3$, and $4$, and in the $XY$ spin glass ($XY$SG) with
Gaussian interactions in dimension $4$.  In all systems with a nonzero
ordering temperature, $x(T)$ is strongly temperature dependent below as
well as above $T_{\rm c}$. The effective exponent initially decreases
progressively as $T$ increases from zero; it passes through a minimum at a
temperature $T_{\rm min}$ close to $T_{\rm c}$, and from then on
increases sharply. The data for the mean-field SK spin glass and for the
finite-dimensional systems follow strikingly similar patterns.
We associate the observed minimum with the critical behavior in 
Eq.~(\ref{eq_crit}) which potentially provides a powerful criterion for
identifying $T_{\rm c}$ in spin glasses from purely energetic measurements. 

The value of the stiffness exponent in the $d = 3$ gauge glass has
been source of controversy: results have clustered either close to
$\theta_{\rm DW} \approx 0.0$
(Refs.~\onlinecite{reger:91,gingras:92,kosterlitz:97,maucourt:97}) or close to
$\theta_{\rm DW} \approx 0.27$
(Refs.~\onlinecite{akino:02,cieplak:92,moore:94,kosterlitz:98,katzgraber:02a}). 
This situation has been analyzed by Akino and 
Kosterlitz,\cite{kosterlitz:98,akino:02} who show that following the boundary
conditions imposed, either a ``best twist'' (BT) value (near $0.27$) or a
``random twist'' (RT) value (near zero) of $\theta_{\rm DW}$ is obtained.
They associate the BT value with domain walls, but they say ``We do not
understand what, if anything, $\theta_{\rm RT}$ means...,'' implying that
this $\theta$ could be nonphysical and simply an artifact arising from an
inappropriate choice of boundary conditions. If interpreted in terms of a
domain-wall stiffness exponent, our $T=0$ estimate is compatible with 
$\theta_{\rm DW} \approx 0$.

In the canonical [Ising,$XY$] Edwards-Anderson spin glass [Ising,$XY$] spins
on a hypercubic lattice of size $L$ interact with their nearest neighbors
through random interactions whose strengths follow a Gaussian distribution
with zero mean and standard deviation unity\cite{binder:86}:
\begin{equation}
{\cal H} = - \sum_{\langle i, j\rangle} J_{ij}S_i S_j \; .
\label{hamiltonian_ea}
\end{equation}
Periodic boundary conditions are applied.  The mean-field limit system is
the SK model. In the gauge glass,\cite{olson:00} $XY$ spins on a
hypercubic lattice of size $L$ interact through the Hamiltonian
\begin{equation}
{\cal H} = -J \sum_{\langle i, j\rangle} \cos(\phi_i - \phi_j - A_{ij}),
\label{hamiltonian_gg}
\end{equation}
where the sum ranges over nearest neighbors. $\phi_i$ represent the
angles of the spins and $A_{ij}$ are quenched random variables (gauge
fields) uniformly distributed between $[0,2\pi]$ with the constraint that
$A_{ij} = - A_{ji}$. $J = 1$ and periodic boundary
conditions are applied.

In the present series of simulations, samples are equilibrated using the
parallel tempering Monte Carlo method.\cite{hukushima:96,marinari:98b}
The sizes studied are
$N=36, 64, 100, 121, 144$, and $196$ for the SK model,
$L = 3$ -- $6$ in the $d = 4$ ISG,
$L = 2$ -- $5$ in the $d = 4$ GG,
$L = 3$ -- $5$ in the $d = 4$ $XY$SG,
$L = 3$ -- $6$, $8$ in the $d = 3$ ISG,
$L = 2$ -- $6$, $8$ in the $d = 3$ GG,
$L = 3$, $4$, $6$, $8$, $10$, and $12$ in the $d = 2$ ISG, and
$L = 6$, $8$, $12$, and $16$ in the $d = 2$ GG.

The equilibrium energies $e(L,T)$ for a given system are averaged
over at least $10^3$ disorder realizations for the largest system sizes.
Due to small differences in the random interactions, there are
sample to sample fluctuations at each size. At each temperature, the data
for each system are fitted using Eq.~(\ref{eq_energy}), with
$e_{\infty}(T)$, $a(T)$, and $x(T)$ as temperature-dependent fitting
parameters. Note that for each value of $L$, exactly the same set of
samples is studied over the whole range of temperatures.

\begin{figure}
\centerline{\epsfxsize=7.5cm \epsfbox{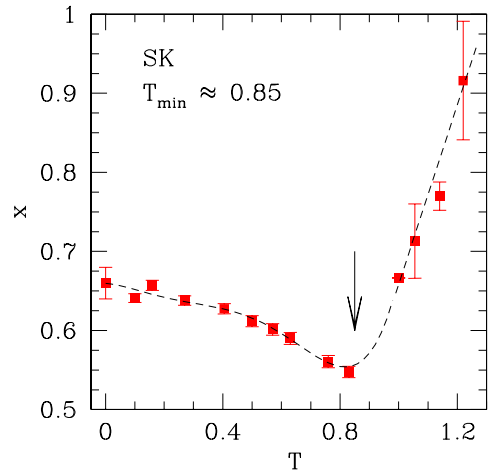}}
\vspace{-1.0cm}
\caption{
Data for $x(T)$ for the SK model. The data show a minimum at $T_{\rm min}
\approx 0.85$ and $x$ extrapolates to the mean-field value for $T \rightarrow
0$, $2/3$. The point at $T = T_{\rm c} = 1$ is the exact 
result (Ref.~\onlinecite{parisi:93}). In this and all subsequent figures the 
dashed line is a guide to the eye and the arrow marks the position of 
$T_{\rm min}$.
}
\label{fig_sk_nd}
\end{figure}

We first discuss the mean-field SK model and then the
finite-dimension models in order of decreasing dimension.
In the SK model, $x(T) = 2/3$ at the ordering temperature 
$T_{\rm c} = 1$,\cite{parisi:93} in agreement with 
Eq.~(\ref{eq_crit}). For $T > T_{\rm c}$ the
power law [see Eq.~(\ref{eq_energy})] is an approximation to a sum of
terms in $N^{-k}$ with $k \in {\mathbb N}$. We fit $e(N,T)$ assuming that
the power law with exponent $x(T)$ is valid at all temperatures below
$T_{\rm c}=1$ and is a useful approximation above. The infinite-size limit
energies $e_\infty(T)$ have been established numerically to high 
precision,\cite{crisanti:02} and we use these values as input in our fits. In
Fig.~\ref{fig_sk_nd}, it can be seen that $x(T)$ is temperature dependent
and passes through a minimum at $T \approx 0.85$. At $T=0$, $x(T)$ again
tends to a value very close to $2/3$, consistent with a directly measured
zero-temperature estimate.\cite{bouchaud:02}

\begin{figure}
\centerline{\epsfxsize=7.5cm \epsfbox{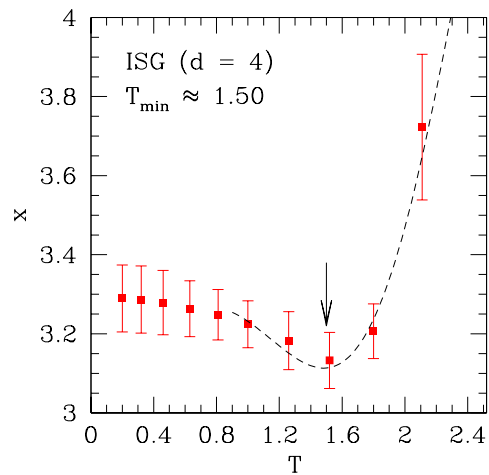}}
\vspace{-1.0cm}
\caption{
Data for $x(T)$ for the $d = 4$ ISG. The data show a minimum at $T_{\rm min}
\approx 1.5$, close to standard estimates of $T_{\rm c}$. We find
$\theta_{\rm E}(T\rightarrow 0) = 0.71 \pm 0.08$ .
}
\label{fig_ea_4d}
\end{figure}

For the $d = 4$ ISG (Fig.~\ref{fig_ea_4d}), $\theta_{\rm E}(0) = 0.71 \pm
0.08$ , in agreement with estimates for the $T = 0$ 
domain-wall stiffness of the bimodal $d = 4$ ISG: $\theta_{\rm DW} =
0.65\pm 0.04$ (Ref.~\onlinecite{hartmann:99}) and 
$0.82 \pm 0.06$ (Ref.~\onlinecite{hukushima:99}).
However, as the two estimates are fairly different from each
other, a more stringent comparison must await a definitive estimate for
$\theta_{\rm DW}(0)$. There is a minimum in $x(T)$ at $T \approx 1.5$, a
temperature lower than $T_{\rm c}=1.80 \pm 0.02$ (Refs.~\onlinecite{parisi:96}
and \onlinecite{campbell:00}).
We find $x(T_{\rm c}) = 3.10 \pm 0.05$, which is higher than 
$d - 1/\nu = 2.8 \pm 0.1$.

For the GG in dimension $4$, our data only extend down to $T = 0.7$; $x(T)$
has a minimum at $T \approx 0.85$, slightly below the ordering temperature
($T_{\rm c} = 0.89 \pm 0.01$, Ref.~\onlinecite{katzgraber:03a}). 
By extrapolation, we
estimate $\theta_{\rm E}(0) = 0.54 \pm 0.05$ and $x(T_{\rm c})=3.42
\pm 0.02$, which, with $\nu = 0.70 \pm 0.1$,\cite{katzgraber:03b} is
distinctly larger than $d - 1/\nu = 2.6 \pm 0.2$.

\begin{figure}
\centerline{\epsfxsize=7.5cm \epsfbox{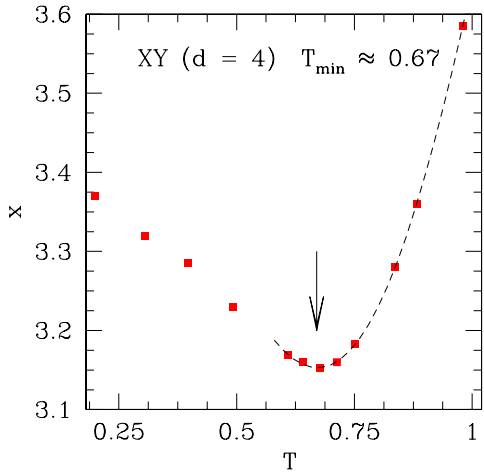}}
\vspace{-1.0cm}
\caption{
Data for $x(T)$ for the four-dimensional $XY$SG. As we have the same number
of system sizes as fitting parameters, we cannot estimate an error bar to
the fits. The data show a clear minimum at $T_{\rm min} = 0.67 \pm 0.02$.
}
\label{fig_xy_4d}
\end{figure}

We have also analyzed data on the four-dimensional $XY$SG with Gaussian
interactions (see Fig.~\ref{fig_xy_4d}).
The correction exponent $x(T)$ behaves similar to the
other $d = 4$ cases, with $\theta_{\rm E}(0)= 0.60\pm 0.05$ and a well
defined minimum at $T_{\rm min}=0.67 \pm 0.02$. We are not aware of
measurements of $\theta_{\rm DW}(0)$ or of $T_{\rm c}$ for this system,
although $T_{\rm c}\approx 0.95$ has been reported for the
four-dimensional $XY$SG with bimodal interactions.\cite{jain:96}

\begin{figure}
\centerline{\epsfxsize=7.5cm \epsfbox{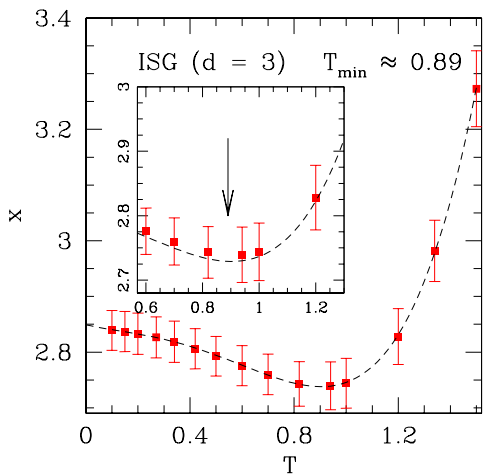}}
\vspace{-1.0cm}
\caption{
Data for $x(T)$ for the Ising spin glass in dimension 3. We see a
minimum which coincides with $T_{\rm c}$ [this is enlarged in detail in the
inset of the figure], $T_{\rm min} \approx 0.89$, $x_{\rm E}(0) \approx
2.85$.
}
\label{fig_ea_3d}
\end{figure}

Figure \ref{fig_ea_3d} shows a similar behavior for the $d = 3$ ISG.
Again, $x(T)$ initially decreases as temperature increases.
The data are consistent with two
temperature points reported for $T=0.7$ and $T=0.8$ by Komori {\em et
al}.~\cite{komori:99}.  The extrapolated low-temperature limiting value is
$\theta_{\rm E}(0)=0.15\pm 0.02$.  This is consistent with a
zero-temperature measurement 
$\theta_{\rm E}(0)=0.135 \pm 0.037$.\cite{bouchaud:02}
The values from the three independent studies taken
together suggest a zero-temperature limiting value $\theta_{\rm E}(0)$
that is significantly lower than the directly measured three-dimensional
domain-wall exponent value 
$\theta_{\rm DW}(0)=0.19\pm 0.01$.\cite{bray:87,hartmann:99}
There is a minimum in $x(T)$ at a temperature
$T_{\rm min} \approx 0.89$ which, in this case, agrees within the quoted
errors to $T_{\rm c}$ estimated independently by other 
techniques,\cite{parisi:96,mari:99} $T_{\rm c} =0.94 \pm 0.03$.
Using $\nu = 1.65 \pm 0.1$ (Refs.~\onlinecite{parisi:96} and 
\onlinecite{mari:99}), $d - 1/\nu = 2.4 \pm
0.05$, a value below $x(T_{\rm c}) \approx 2.75$.

\begin{figure}
\centerline{\epsfxsize=7.5cm \epsfbox{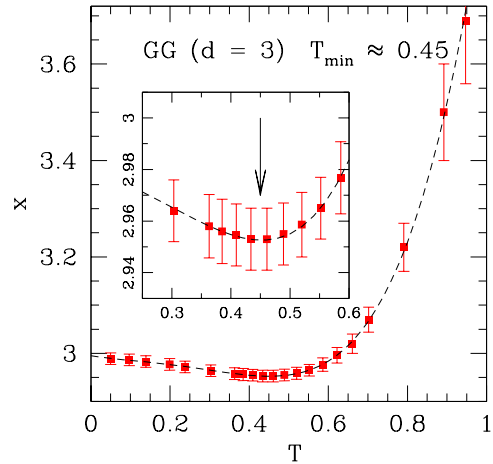}}
\vspace{-1.0cm}
\caption{
Data for the three-dimensional gauge glass $x(T)$. The data show a minimum
at $T_{\rm min} \approx 0.45$. In addition, $x(0) \approx 2.99$. In the
inset the data are enlarged around the minimum.
}
\label{fig_gg_3d}
\end{figure}

Figure \ref{fig_gg_3d} shows $x$ vs $T$ for the $d = 3$ GG. Here, the 
low-temperature limit corresponds to $\theta_{\rm E} = 0.010(12)$. As in the
other cases, $x(T)$ decreases with increasing $T$. There is a
minimum in $x(T)$ at $T \approx 0.45$, which again, within the error bars,
agrees with the ordering temperature $T_{\rm c}$ estimated
from other methods.\cite{olson:00,katzgraber:02a}
We find $d - 1/\nu = 2.28 \pm 0.03$ 
($\nu=1.39 \pm 0.03$, Ref.~\onlinecite{katzgraber:03b}), 
whereas the measured value is
much larger, $x(T_{\rm c}) = 2.95 \pm 0.01$.

The two-dimensional ISG with Gaussian random interactions
does not order above zero temperature.\cite{hartmann:01a}
Our data show that $x(T)$ tends to $2.37 \pm 0.06$ at
zero temperature, consistent with the value of $2.35 \pm 0.02$ observed in
Ref.~\onlinecite{bouchaud:02}. Both estimates suggest that $\theta_{\rm E}(0)$
is slightly more negative than the accurately measured $\theta_{\rm DW}= -
0.28 \pm0.01$.\cite{rieger:96,hartmann:01a,carter:02} 
$x(T)$ shows a
shallow minimum at $T \approx 0.4$, unrelated to any ordering temperature.

In the $d = 2$ GG, for which $T_{\rm c} = 0$,\cite{katzgraber:03a} $x(T)$
increases steadily as $T$ rises, from a value $x(0.13)=2.40 \pm 0.02$ at the
lowest measuring temperature $T=0.13$ corresponding to $\theta_{\rm E}(0)
\approx -0.40 \pm 0.02$ in agreement with results from
Ref.~\onlinecite{katzgraber:02a}. Note also that the above estimate of
$x(T \rightarrow T_{\rm c} = 0)$ agrees with the zero-$T_{\rm c}$ scaling 
of Eq.~(\ref{eq_crit}).

We can review the estimates for $\theta_{\rm E}(0)$ in the GG. In
dimension 2, the estimated $\theta_{\rm E}(0) = -0.40 \pm 0.02$ could be
consistent with either the ``best twist'' $\theta_{\rm BT}$, which is
$-0.39 \pm 0.03$,\cite{kosterlitz:98,katzgraber:02a} or with the ``random
twist''$\theta_{\rm RT}$, $-0.45 \pm 0.015$.\cite{kosterlitz:98} 
For the three-dimensional GG, however, the low-temperature limit $\theta_{\rm
E}(0) = 0.010 \pm 0.012$ is in agreement within the error bars with the
$\theta_{\rm RT}$ estimate by Akino and Kosterlitz, $0.05\pm 0.05$
(Ref.~\onlinecite{akino:02}) but is completely different from the domain-wall
$\theta_{\rm BT}(0)$ estimated to be 
$0.27 \pm 0.01$.\cite{katzgraber:02a,akino:02}
For the GG in dimension $4$, our data do
not go below $T=0.7$; we estimate $\theta_{\rm E}(0) = 0.54 \pm 0.05$. By
extrapolation from dimensions $1$ (Ref.~\onlinecite{bray:84}), 
$2$ and $3$, we estimate
for the four-dimensional GG $\theta_{\rm BT} \approx 0.9 \pm 0.1$. We
conclude that in the GG $\theta_{\rm E}(0)$, defined unambiguously through
the effect of periodic boundary conditions on the energy, can be very
different from the domain-wall $\theta_{\rm BT}(0)$; it may well be
possible to identify it with $\theta_{\rm RT}(0)$ as defined by Akino
and Kosterlitz.\cite{akino:02}
Regardless, the size effect provides an
operational physical realization of a bona fide $\theta$ quite
distinct from the ``best twist'' domain-wall stiffness exponent
$\theta_{\rm BT}(0)$. The strong difference between the domain-wall
exponent $\theta_{\rm BT}(0)$ and the periodic boundary conditions
exponent $\theta_{\rm E}(0)$ seen in $d = 3$ and implied in $d = 4$ is
presumably related to the extra liberty that vector spins (as opposed to
Ising spins) have to reorganize under external constraints. Any
differences there may be between $\theta_{\rm E}(0)$ and $\theta_{\rm
DW}(0)$ in ISG's certainly appear to be less spectacular than in the GG.
However, in both dimension 3 and dimension 2, independent measurements of
$\theta_{\rm E}(0)$ are consistent with each other while the agreement
with $\theta_{\rm DW}(0)$ is poor.

We find that for the models studied, the observed critical
exponent $x(T_{\rm c})$ is systematically higher than its rigorous 
scaling value $d - 1/\nu$. A possible explanation might be the influence
of a ``lattice artifact'' correction\cite{salas:98,ballesteros:99} so that,
for instance, $e(L) - e_{\infty}= aL^{-(d-1/\nu)} + bL^{-d}$.
In the presence of this correction term the data can be represented quite 
accurately by a single effective exponent $x$ with a value between 
$d - 1/\nu$ and $d$ if $b$ is positive. 

In conclusion, we have presented numerical results on the scaling
correction to the internal energy per spin as a function of system size
and temperature in a variety of spin-glass models.  
The $T=0$ finite-size correction has been
linked to domain walls,\cite{newman:98,komori:99,bouchaud:02} and scaling 
predicts $x = d - 1/\nu$ at $T_{\rm c}$. By
definition, the domain-wall stiffness exponent $\theta_{\rm DW}(T)$ drops
to zero at $T_{\rm c}$ (see Refs.~\onlinecite{olson:00},
\onlinecite{katzgraber:02a} and \onlinecite{hukushima:99}) 
and one would expect from the domain-wall picture that 
$\theta_{\rm E}(T)$ should drop in a similar way. 
Based on standard scaling arguments, the effective
stiffness exponent $\theta_{\rm E}(T = T_{\rm c})$ should be equal to $1/\nu$.
In practice, for $0 \le T \le T_{\rm c}$ there is a 
steady enhancement of the effective stiffness exponent 
in all finite-$T_c$ spin-glass systems studied, which can be interpreted as 
a gradual change from domain-wall to critical behavior. Nevertheless,
the temperature-dependence of $\theta_{\rm E}$  is incompatible with
a domain-wall mechanism.
However, the critical dip is never as deep as the scaling theory 
predicts -- the observed effective $\theta_{\rm E}(T_{\rm c})$ is 
systematically smaller than $1/\nu$. A rigorous
relationship between the minimum of $x(T)$, $T_{\rm min}$, and $T_{\rm c}$
could provide a method to determine $T_{\rm c}$ from purely energetic
measurements.

We are extremely grateful to A.~Crisanti and T.~Rizzo for generously
providing us with the tabulated data for the infinite-size SK energies
and to G.~Blatter, A.~K.~Hartmann, and F.~Ricci-Tersenghi
for discussions. We would also like to thank A.~K.~Hartmann for critically
reading the manuscript.

\bibliography{refs}

\end{document}